%% file: misleading.tex
\documentclass{article}

\usepackage{complexity}                
\usepackage{hyperref}                  
\usepackage{fullpage}                  
\usepackage{booktabs}                  
\usepackage{longtable}                 
\usepackage{tikz}                      
\usepackage{wrapfig}                   
\usepackage{placeins}                  
\usepackage{caption}                   
\usepackage{amsmath}                   

\usetikzlibrary{arrows, shadows, calc, positioning}
\usetikzlibrary{decorations.pathreplacing, decorations.pathmorphing}

\definecolor{safelightblue}{rgb}{0.65098, 0.807843, 0.890196}
\definecolor{safedarkblue}{rgb}{0.121569, 0.470588, 0.705882}
\definecolor{safelightorange}{rgb}{0.996078, 0.901961, 0.807843}
\definecolor{safedarkorange}{rgb}{0.54902, 0.176471, 0.015686}

\title{Greedy Graph Colouring is a Misleading Heuristic}

\author{Ciaran McCreesh and Patrick Prosser}

\begin{document}

\maketitle

Let $G = (V, E)$ be a (finite, simple, undirected) graph with vertex set $V$ and edge set $E$. A
\emph{clique} in $G$ is a set of vertices, each of which is adjacent to every other vertex in this
set. The \emph{maximum clique problem} is to find a largest such set in a given graph; this is
\NP-hard \cite{Garey:1990}. We denote the size of a maximum clique by $\omega$.

A \emph{colouring} of $G$ is an assignment of vertices to colours such that no two adjacent vertices
are given the same colour. Determining the minimum number of colours $\chi$ required to colour a
graph is also \NP-hard \cite{Garey:1990}, but greedy algorithms may produce a (non-optimal)
colouring in polynomial time. There are two ``obvious'' quadratic greedy colouring algorithms: one
could give each vertex in turn the first available colour. Alternatively, for each colour in turn,
one could try to give that colour to each vertex in turn. In fact these two algorithms produce the
same result.

\begin{figure}[htb]
    \centering
    \begin{tikzpicture}
        \input{figure-clique}
    \end{tikzpicture}\hspace{3em}\begin{tikzpicture}
        \input{figure-colour}
    \end{tikzpicture}
    \caption{On the left, a graph, with its unique maximum clique $\{1, 3, 6, 8\}$ of size 4
        highlighted. On the right, a greedy colouring of the same graph: first, vertices $\{1, 2, 4,
        7\}$ are coloured in pale cream. Next, vertices $\{3, 5, 9\}$ are coloured light blue, then
        $6$ is coloured in medium blue, and finally $8$ is coloured in dark brown. We can infer that
        any clique of size four must include vertices $6$ and $8$, plus one pale cream vertex and
        one light blue vertex. In this case, the maximum clique number, colour number, and greedy
        colouring number all agree; usually we are not so fortunate.}
    \label{figure:clique-and-colour}
\end{figure}
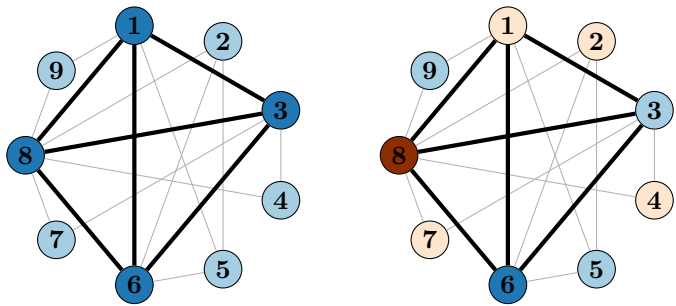

Any colouring of a graph require at least $\omega$ colours (each vertex in a clique must be given a
different colour). Thus a greedy colouring provides the bound part of a branch and bound algorithm
for the maximum clique problem. We illustrate these concepts in
Figure~\ref{figure:clique-and-colour}, and refer to papers by Tomita for algorithms
\cite{Tomita:2007,Tomita:2010}.

\begin{figure}[htb]
    \centering
    \begin{tikzpicture}
        \input{figure-misleading-left}
    \end{tikzpicture}\hspace{3em}\begin{tikzpicture}
        \input{figure-misleading-right}
    \end{tikzpicture}

    \caption{Greedy colouring is a misleading heuristic. On the left, a graph which, when coloured
        greedily in numerical order requires two colours: vertices $\{1, 3, 5, 7\}$ are coloured
        light blue, and vertices $\{2, 4, 6\}$ are coloured medium blue. On the right, the same
        graph with the first vertex removed. Now a greedy colouring in the same order will require
        three colours: first it will colour vertices $2$ and $3$ in light blue, then vertices $4$
        and $6$ in medium blue. The remaining two vertices now require a third colour, very light
        cream. This shows that the bound for a subtree can be worse than the bound for its parent, which
        has implications for parallelism.}
    \label{figure:misleading}
\end{figure}
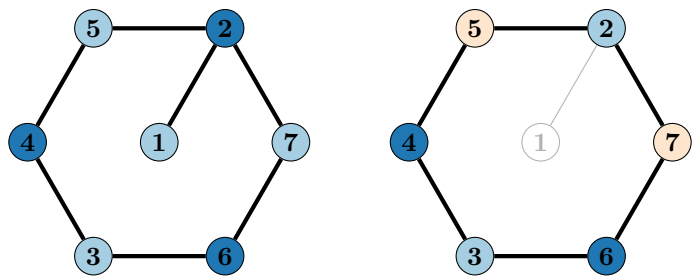

A bound is said to be \emph{misleading} if it is possible that the value of the bound function for a
problem could be better than the value for a subproblem. Unfortunately a greedy colouring is
misleading: sometimes, it can colour a graph $G$ with $k$ colours, but then require $k + 1$ (for
arbitrarily large values of 1) colours to colour an induced subgraph of $G$. We give one example in
Figure~\ref{figure:misleading}.

There are two implications. Firstly, we could potentially improve Tomita-style maximum clique
algorithms by passing the number of colours used in a parent problem as an extra parameter to the
recursive function, and use the minimum of this number and the number of colours as a bound.
Experiments with the sensibly-sized DIMACS graphs show that although situations similar to those in
Figure~\ref{figure:misleading} do occur regularly, in no case does this improved bound help
eliminate any subproblems (i.e.\ search node counts are unchanged).

Secondly, a non-misleading heuristic is one of the necessary conditions for avoiding a slowdown in
parallel branch and bound algorithms \cite{Li:1986,Trienekens:1990,deBruin:1995}. Thus parallel
maximum clique algorithms, such as those by the authors \cite{McCreesh:2013} and Depolli et.\ al.\
\cite{Depolli:2013} should probably be adapted slightly to avoid the possibility of strange
behaviour.

\FloatBarrier

\bibliographystyle{amsalpha}
\bibliography{misleading}

\end{document}

%% file: figure-clique.tex

\newcount \c
\foreach \n in {1, ..., 9}{
    \c=\n
    \multiply\c by -40
    \advance\c by 140
    \ifthenelse{\n = 1 \OR \n = 3 \OR \n = 6 \OR \n = 8}{
        \node[draw, circle, fill=safedarkblue, inner sep=2pt] (N\n) at (\the\c:1.75) {\textbf{\n}};
    }{
        \node[draw, circle, fill=safelightblue, inner sep=2pt] (N\n) at (\the\c:1.75) {\textbf{\n}};
    }
}

\tikzstyle{h} = [ultra thick];
\tikzstyle{l} = [color = black!30!white];
\draw [l] (N1) -- (N5); \draw [l] (N1) -- (N9);
\draw [l] (N2) -- (N5); \draw [l] (N2) -- (N6); \draw [l] (N2) -- (N8);
\draw [l] (N3) -- (N4); \draw [l] (N3) -- (N7);
\draw [l] (N4) -- (N8);
\draw [l] (N5) -- (N6);
\draw [l] (N7) -- (N8);
\draw [l] (N8) -- (N9);

\draw [h] (N1) -- (N3);
\draw [h] (N6) -- (N8);
\draw [h] (N1) -- (N6);
\draw [h] (N1) -- (N8);
\draw [h] (N3) -- (N6);
\draw [h] (N3) -- (N8);

%% file: figure-colour.tex

\newcount \c
\foreach \n in {1, ..., 9}{
    \c=\n
    \multiply\c by -40
    \advance\c by 140
    \ifthenelse{\n = 1 \OR \n = 2 \OR \n = 4 \OR \n = 7}{
        \node[draw, circle, fill=safelightorange, inner sep=2pt] (N\n) at (\the\c:1.75) {\textbf{\n}};
    }{
        \ifthenelse{\n = 3 \OR \n = 5 \OR \n = 4 \OR \n = 7 \OR \n = 9}{
            \node[draw, circle, fill=safelightblue, inner sep=2pt] (N\n) at (\the\c:1.75) {\textbf{\n}};
        }{
            \ifthenelse{\n = 6}{
                \node[draw, circle, fill=safedarkblue, inner sep=2pt] (N\n) at (\the\c:1.75) {\textbf{\n}};
            }{
                \node[draw, circle, fill=safedarkorange, inner sep=2pt] (N\n) at (\the\c:1.75) {\textbf{\n}};
            }
        }
    }
}

\tikzstyle{h} = [ultra thick];
\tikzstyle{l} = [color = black!30!white];
\draw [l] (N1) -- (N5); \draw [l] (N1) -- (N9);
\draw [l] (N2) -- (N5); \draw [l] (N2) -- (N6); \draw [l] (N2) -- (N8);
\draw [l] (N3) -- (N4); \draw [l] (N3) -- (N7);
\draw [l] (N4) -- (N8);
\draw [l] (N5) -- (N6);
\draw [l] (N7) -- (N8);
\draw [l] (N8) -- (N9);

\draw [h] (N1) -- (N3);
\draw [h] (N6) -- (N8);
\draw [h] (N1) -- (N6);
\draw [h] (N1) -- (N8);
\draw [h] (N3) -- (N6);
\draw [h] (N3) -- (N8);

%% file: figure-misleading-left.tex

\node[draw, circle, fill=safelightblue, inner sep=2pt] (N1)            {\textbf{1}};
\node[draw, circle, fill=safedarkblue, inner sep=2pt] (N2) at (60:1.75)  {\textbf{2}};
\node[draw, circle, fill=safelightblue, inner sep=2pt] (N3) at (120:1.75) {\textbf{5}};
\node[draw, circle, fill=safedarkblue, inner sep=2pt] (N4) at (180:1.75) {\textbf{4}};
\node[draw, circle, fill=safelightblue, inner sep=2pt] (N5) at (240:1.75) {\textbf{3}};
\node[draw, circle, fill=safedarkblue, inner sep=2pt] (N6) at (300:1.75) {\textbf{6}};
\node[draw, circle, fill=safelightblue, inner sep=2pt] (N7) at (360:1.75) {\textbf{7}};

\tikzstyle{h} = [ultra thick];
\draw [h] (N1) -- (N2);
\draw [h] (N2) -- (N3);
\draw [h] (N3) -- (N4);
\draw [h] (N4) -- (N5);
\draw [h] (N5) -- (N6);
\draw [h] (N6) -- (N7);
\draw [h] (N7) -- (N2);

%% file: figure-misleading-right.tex

\node[draw, circle, color=black!30!white, text=black!30!white, inner sep=2pt] (N1) {\textbf{1}};

\node[draw, circle, fill=safelightblue,   inner sep=2pt] (N2) at (60:1.75)  {\textbf{2}};
\node[draw, circle, fill=safelightorange, inner sep=2pt] (N3) at (120:1.75) {\textbf{5}};
\node[draw, circle, fill=safedarkblue,    inner sep=2pt] (N4) at (180:1.75) {\textbf{4}};
\node[draw, circle, fill=safelightblue,   inner sep=2pt] (N5) at (240:1.75) {\textbf{3}};
\node[draw, circle, fill=safedarkblue,    inner sep=2pt] (N6) at (300:1.75) {\textbf{6}};
\node[draw, circle, fill=safelightorange, inner sep=2pt] (N7) at (360:1.75) {\textbf{7}};

\tikzstyle{h} = [ultra thick];
\tikzstyle{l} = [color = black!30!white];
\draw [l] (N1) -- (N2);

\draw [h] (N2) -- (N3);
\draw [h] (N3) -- (N4);
\draw [h] (N4) -- (N5);
\draw [h] (N5) -- (N6);
\draw [h] (N6) -- (N7);
\draw [h] (N7) -- (N2);